\documentclass{revtex4}
\usepackage{graphicx}
\usepackage{amsmath}

\begin{document}

\title{Flux tube dressed with color electric $E^a_{\rho,\varphi}$ and 
magnetic $H^a_{\rho,\varphi}$ fields}
\author{Vladimir Dzhunushaliev}
\email{dzhun@hotmail.kg}
\affiliation{Dept. Phys. and Microel. Engineer., Kyrgyz-Russian
Slavic University, Bishkek, Kievskaya Str. 44, 720000, Kyrgyz
Republic}

\date{\today}

\begin{abstract}
The nonperturbatuve analytical calculations in quantum field theory are 
presented. On the first step the Nielsen - Olesen flux tube in the SU(2) 
Yang - Mills - Higgs theory dressed by color electric $E^a_{\rho,\varphi}$ 
and magnetic $H^a_{\rho,\varphi}$ fields is derived. On the next step it is 
shown that this flux tube can be considered as a pure nonperturbative 
quantum object in the SU(3) gauge theory. 
\end{abstract}

\maketitle

\section{introduction}

The confinement problem in quantum chromodynamics is a manifestation 
of problems in the quantum field theory with strong interactions. Let us 
compare this situation with the situation in the classical field theory. If a 
classical theory is linear we know the answer for any question: 
for example, in classical electrodynamics 
without currents and charges any solution is the superposition of simple 
harmonic waves. If such theory is nonlinear we have big problems : for example, 
in Yang -Mills theory we have not any general algorithm for taking solutions and 
in this case the solutions like instantons, monopoles are big success by 
the research. This situation 
becomes deeper in quantum field theory: we can calculate everything in 
QED but the interaction between quark and antiquark is unresolved problem 
for us. The reason is that an algebra of quantized strongly interacting fields 
is much more complicated and richer the algebra of linear fields. 
\par 
The complexity of confinement problem lies in the fact that it is very difficult 
to derive a tube filled with the color longitudinal electric field $E^a_z$. 
In a dual theories such tubes filled with $H_z$ exist: it is well known 
the Nielsen - Olesen flux tube in U(1) gauge theory interacting with Higgs 
scalar field \cite{no} and BPS flux tubes, see for example \cite{kneipp}. 
It is clear also why the flux tube can not be obtained by 
perturbative technique in quantum field theory: on the perturbative 
consideration level any quantized field distribution is a cloud of quanta. 
It is obvious that such field configuration as the flux tube can not be 
the cloud of particles moving with the speed of light. It allows us to say 
that the quantum field theory with strongly interacting fields strongly 
differs from the theories with a small coupling constant. Therefore the 
problem  of confinement is obvious : we have to quantize not a set of 
oscillators but a field distribution as the whole. Probably such quantization 
procedure is similar to the art: just as in the classical theory - finding 
such solutions as black holes, monopoles, instantons and so on is the art 
of researcher. In both cases we do not have any procedure (algorithm) for 
taking classical/quantum field distributions describing one or another 
physical situation. 
\par 
In section II we will consider a cylindrically symmetric solution in the SU(2) 
Yang - Mills - Higgs theory with broken gauge symmetry. The derived solution 
is the Nielsen - Olesen flux tube dressed with color electric 
$E_{\rho,\varphi}$ and magnetic $H_{\rho,\varphi}$ fields. In section III 
we present arguments that this dressed flux tube is a pure quantum object 
in the SU(3) gauge theory. 

\section{Dressed flux tube}

In this paper we assume that in the SU(2) gauge theory exists such quantum 
phenomenon as gauge symmetry breakdown. At present this phenomenon can not 
be calculated on the nonperturbative level and we only suppose that it takes 
place. Then the field equations with mass term which breaks the gauge symmetry 
are 
\begin{eqnarray}
  \mathcal{D}_\nu F^{a\mu\nu} &=& g \epsilon^{abc} \phi^b 
  \mathcal{D}^\mu \phi^c + m^2(a) A^{a\mu} , 
\label{sec1-10}\\
  \mathcal{D}_\mu \mathcal{D}^\mu \phi^a &=& -\lambda \phi^a 
  \left(
  \phi^b \phi^b - \phi^2_\infty
  \right) 
\label{sec1-20}  
\end{eqnarray} 
where $ F^a_{\mu\nu} = \partial_\mu A^a_\nu - \partial_\nu A^a_\mu + 
g\epsilon^{abc} A^b_\mu A^c_\nu$; $A^a_\mu$ is the SU(2) gauge potential;
$a,b,c = 1,2,3$ are color indices; $\mu,\nu$ are Lorentz indices; 
$\phi^a$ is the Higgs field; $g, \lambda$ some constants; $m(a)$ is 
the gauge breaking mass term, in the next section we will discuss how it can 
erase in the SU(3) quantum theory.
\par
\begin{figure}[h]
  \begin{center}
    \fbox{
    \includegraphics[height=5cm,width=5cm]{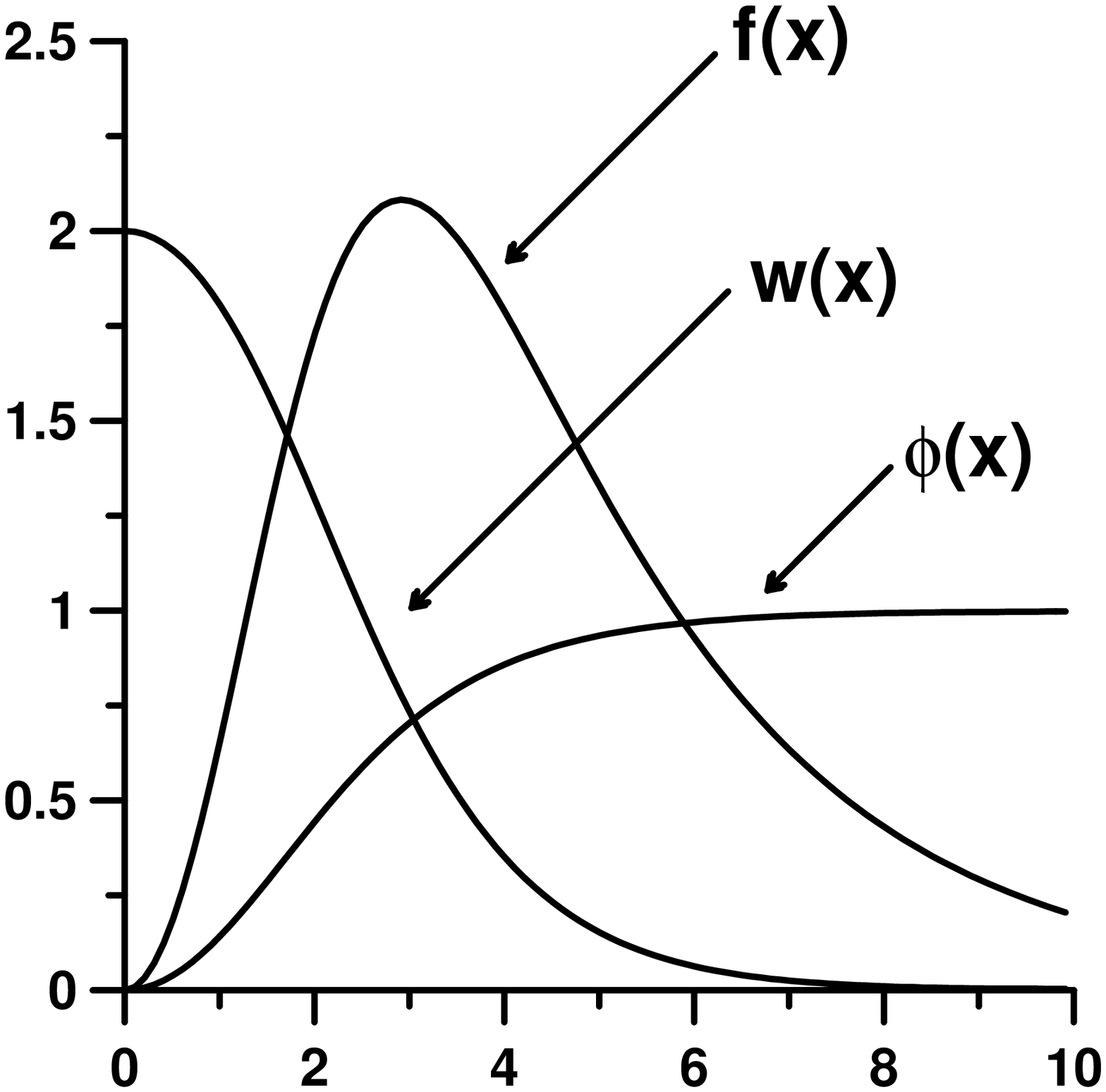}}
    \caption{The functions $f(x)$, $w(x)$ and $\phi(x)$.}
    \label{fig1}
  \end{center}
\end{figure}

 
The solution we search in the following form 
\begin{equation}\label{sec1-30}
    A^1_t = f(\rho) \; , \;
    A^1_z = v(\rho) \; , \;
    A^3_\varphi = w(\rho) \; , \;
    \phi^2 = \phi(\rho)
\end{equation}
where $\rho, z , \varphi$ are the cylindrical coordinates. The substitution 
in \eqref{sec1-10}, \eqref{sec1-20} equations gives us 
\begin{eqnarray}
  f'' + \frac{f'}{x} &=& f
  \left(  
  \phi^2 + \frac{w^2}{x^2} - m^2
  \right) ,\\
\label{sec1-40}
  v'' + \frac{v'}{x} &=& v 
  \left(  
  \phi^2 + \frac{w^2}{x^2} - m^2
  \right) ,\\
\label{sec1-50}
  w'' - \frac{w'}{x} &=& 
  w\left(
  -f^2 + v^2 + \phi^2
  \right), \\
\label{sec1-60}  
  \phi'' + \frac{\phi'}{x} &=& 
  \phi 
  \left(
  -f^2 + v^2 + \frac{w^2}{x^2}
  \right) + \lambda \phi \left( \phi^2 - 1 \right)
\label{sec1-70}
\end{eqnarray}  
here we have introduced the dimensionless coordinate $x = \rho \phi_\infty$; 
redefined $f/\phi_\infty \rightarrow f$, $v/\phi_\infty \rightarrow v$, 
$w/\phi_\infty \rightarrow w$, $\phi/\phi_\infty \rightarrow \phi$ and 
$m(1) = m(2) = m, m(3) = 0$. We will search the solution in the simplest form 
$f = v$. Then we have the following splitted equations set 
\begin{eqnarray}
    -\left( f'' + \frac{f'}{x} \right) + 
    f V_{eff} &=& -\left( 1 - m^2 \right) f ,
\label{sec1-80}\\
    V_{eff} &=& \phi^2 + \frac{w^2}{x^2} - 1
\label{sec1-90}
\end{eqnarray}
and 
\begin{eqnarray}
  w'' - \frac{w'}{x} &=& w \phi^2 ,
  \label{sec1-100} \\
  \phi'' + \frac{\phi'}{x} &=& \phi w^2 + 
  \lambda \phi \left( \phi^2 - 1 \right) .
  \label{sec1-110}
\end{eqnarray} 
We have very interesting equations set: the first equation \eqref{sec1-80} 
is the Schr\"odinger equation with the potential $V_{eff}$ \eqref{sec1-90} 
and ''wave function`` $f(x)$, the second \eqref{sec1-100} and third \eqref{sec1-110} 
equations describe the well known Nielsen - Olesen flux tube filled 
with $H^2_z$ \textit{color} magnetic field. The solution for Nielsen - Olesen flux tube 
is presented on Fig.\ref{fig1} (functions $w(x)$ and $\phi(x)$). 
\par
\begin{figure}[h]
  \begin{center}
    \fbox{
    \includegraphics[height=5cm,width=5cm]{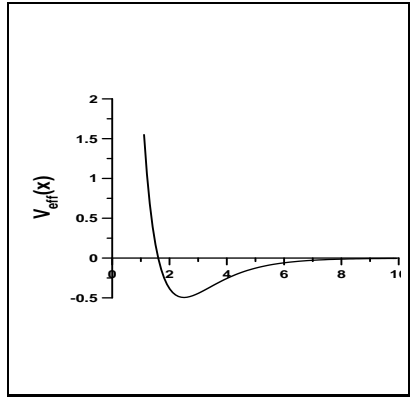}}
    \caption{The potential $V_{eff}(x)$.}
    \label{fig2}
  \end{center}
\end{figure}

This solution has the following behavior at the center of tube 
\begin{eqnarray}
  w(x) &=& 2 + ax^2 + \cdots , 
  \label{sec1-120} \\
  \phi(x) &=& b x^2 +  \cdots 
  \label{sec1-130} 
\end{eqnarray}  
with $a \approx -0.19527$, $b \approx 0.15400139$, $\lambda = 0.25$ 
\cite{Obukhov:1996ry} and $m \approx 0.9402$. At the infinity 
\begin{eqnarray}
  w(x) &\approx& w_0 \sqrt x e^{-x} ,
  \label{sec1-140} \\
  \phi(x) &\approx& 1 - \phi_0 \frac{e^{-x}}{\sqrt x} 
  \label{sec1-150} 
\end{eqnarray}  
here $w_0$ and $\phi_0$ are some constants. This solution gives us 
the potential for the Schr\"odinger equation \eqref{sec1-80} 
presented on the Fig. \ref{fig2}. Immediately we see that the potential has a 
minimum and consequently the corresponding equation \eqref{sec1-80} can have 
a discrete energy spectrum with appropriate wave functions. The numerical 
investigation is presented on Fig. \ref{fig1} (``wave function'' $f(x)$) and 
the corresponding ``energy level'' is $1 - m^2 \approx 0.9402$. 
\par
\begin{figure}[h]
  \begin{minipage}[t]{.45\linewidth}
  \begin{center}
    \fbox{
    \includegraphics[height=5cm,width=5cm]{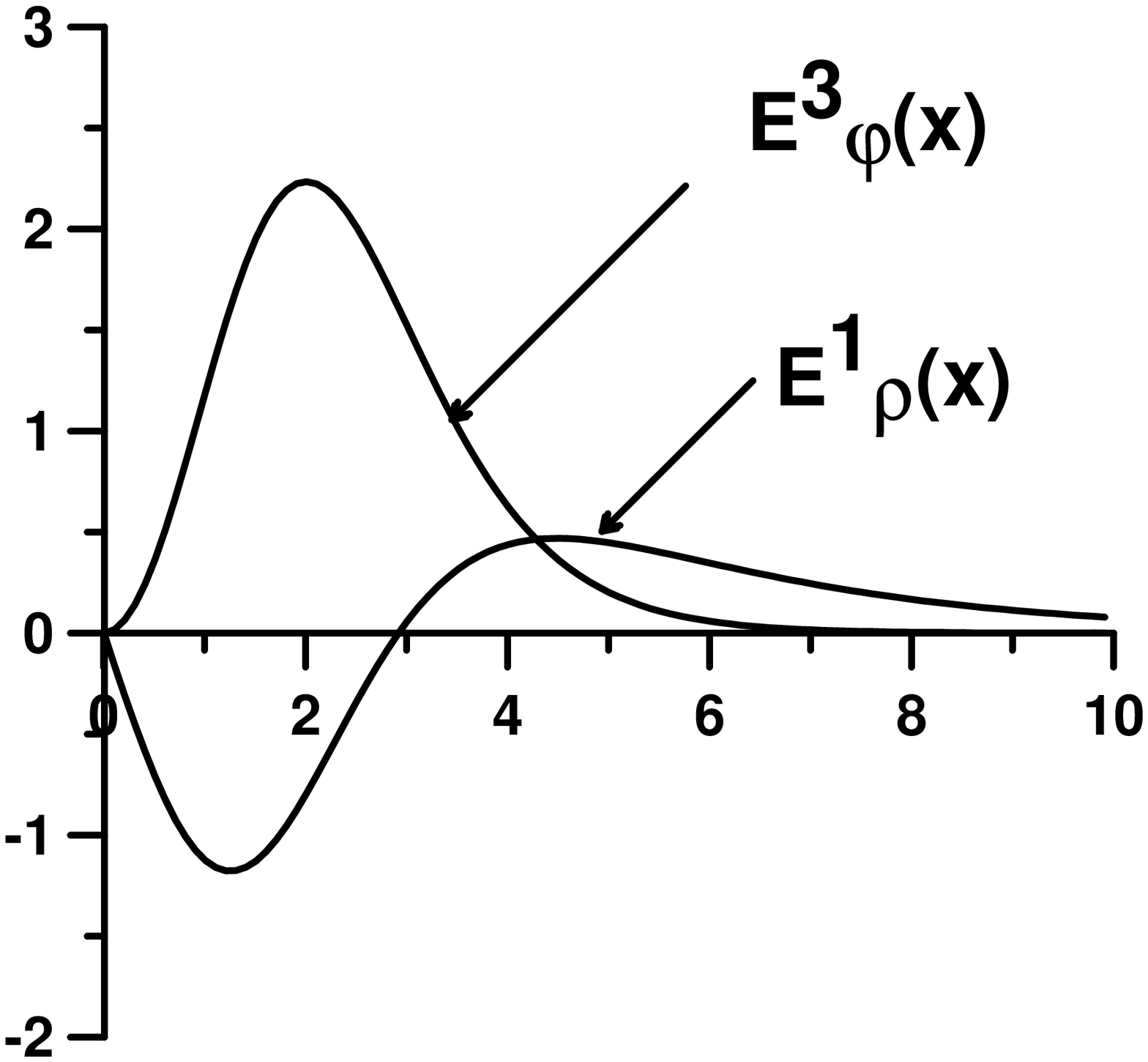}}
    \caption{The electric $E^1_\rho , E^3_\phi$ fields.}
    \label{fig3}
  \end{center}
  \end{minipage}\hfill
  \begin{minipage}[t]{.45\linewidth}
  \begin{center}
    \fbox{
    \includegraphics[height=5cm,width=5cm]{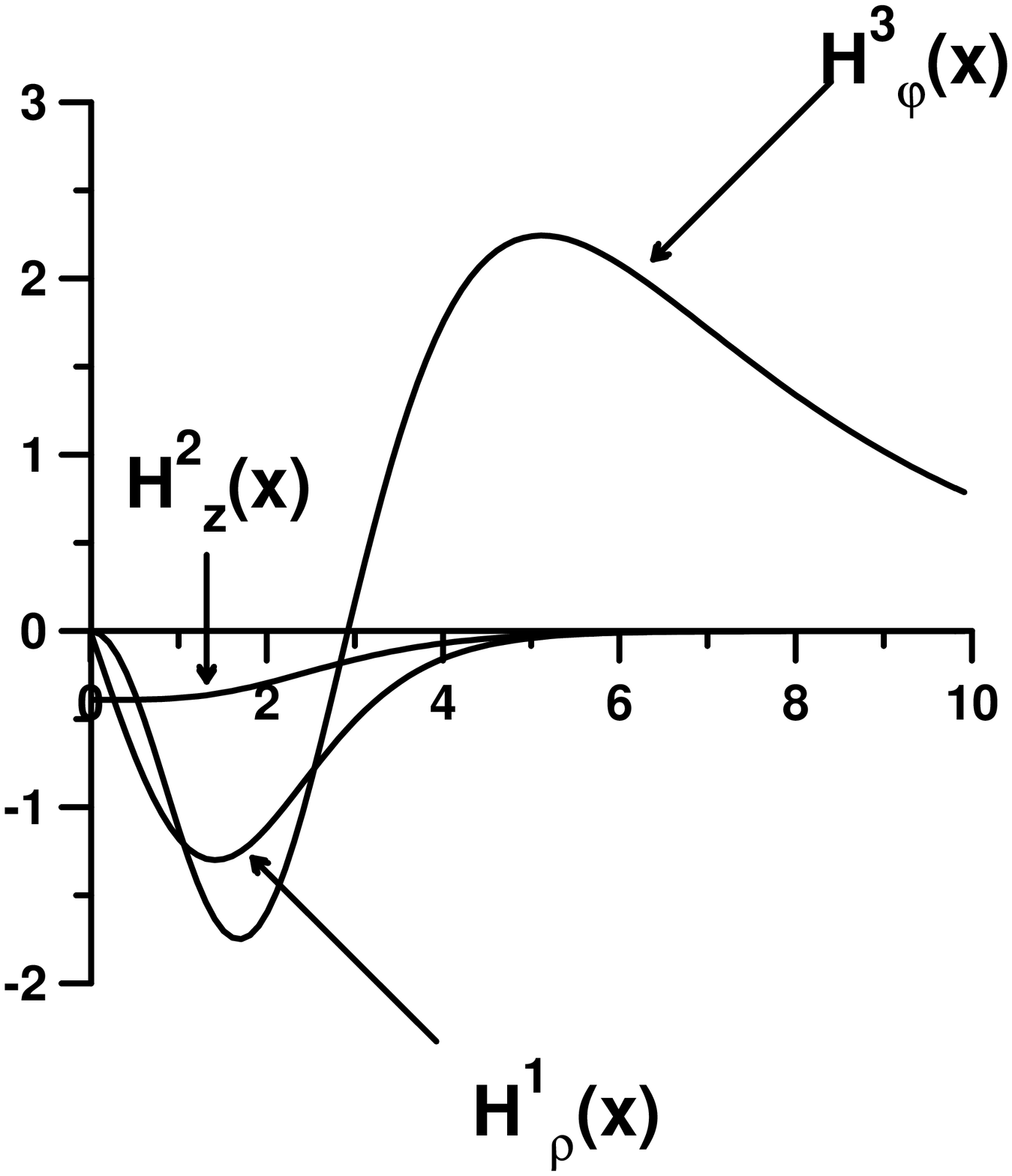}}
    \caption{The magnetic $H^1_\rho , H^3_\phi , H^2_z$ fields.}
    \label{fig4}
  \end{center}
  \end{minipage}
\end{figure}

 
Thus we have derived the gauge potentials $f(x), w(x)$ and $\phi(x)$ 
presented on Fig.\ref{fig1} 
and color electric and magnetic fields shown on Fig's.\ref{fig3},\ref{fig4}. 
We see that it is the Nielsen - Olesen flux tube filled with the longitudinal 
$H^2_z$ magnetic field and dressed with the radial $E^1_\rho, H^1_\rho$ and 
azimuthal $E^3_\varphi, H^3_\varphi$ fields:  
\begin{eqnarray}
  E^3_\varphi = F^3_{t\varphi} = &=& fw ,
  \label{sec1-160} \\
  E^1_\rho = F^1_{t\rho} &=& -f' , 
  \label{sec1-170} \\
  H^1_\varphi = \epsilon_{\varphi z \rho} F^{1 z\rho} &=& -\rho f' , 
  \label{sec1-180} \\
  H^3_\rho = \epsilon_{\rho z \varphi} F^{3 z\varphi} &=& 
  -\frac{1}{\rho} fw 
  \label{sec1-190} \\
  H^2_z &=& \frac{1}{\rho} w' .
  \label{sec1-200}
\end{eqnarray}  
\par
\begin{figure}[h]
  \begin{minipage}[t]{.5\linewidth}
  \begin{center}
    \fbox{
    \includegraphics[height=5cm,width=5cm]{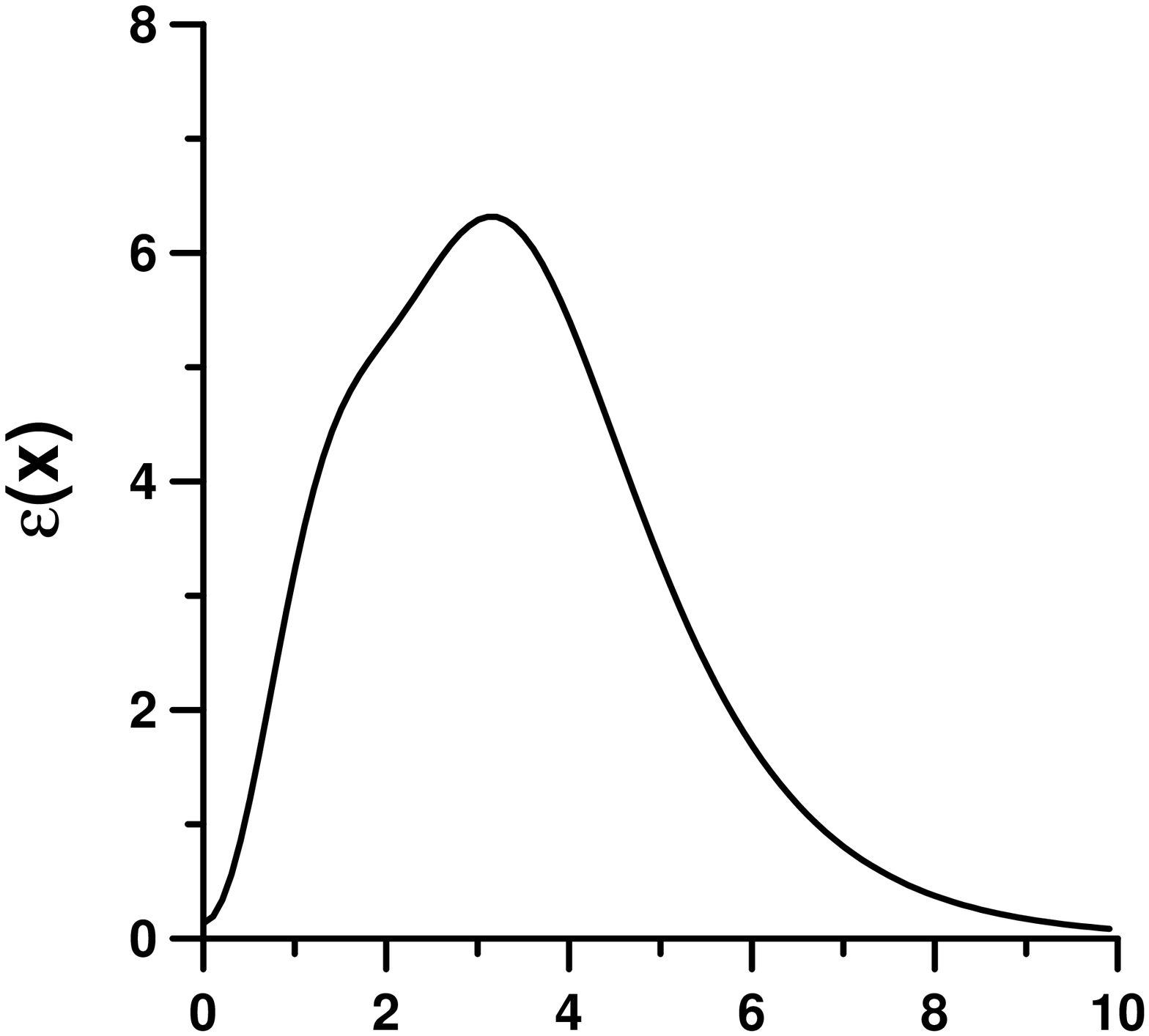}}
    \caption{The energy density $\varepsilon(x)$.}
    \label{fig5}
  \end{center}
  \end{minipage}
  \begin{minipage}[t]{.5\linewidth}
  \begin{center}
    \fbox{
    \includegraphics[height=5cm,width=5cm]{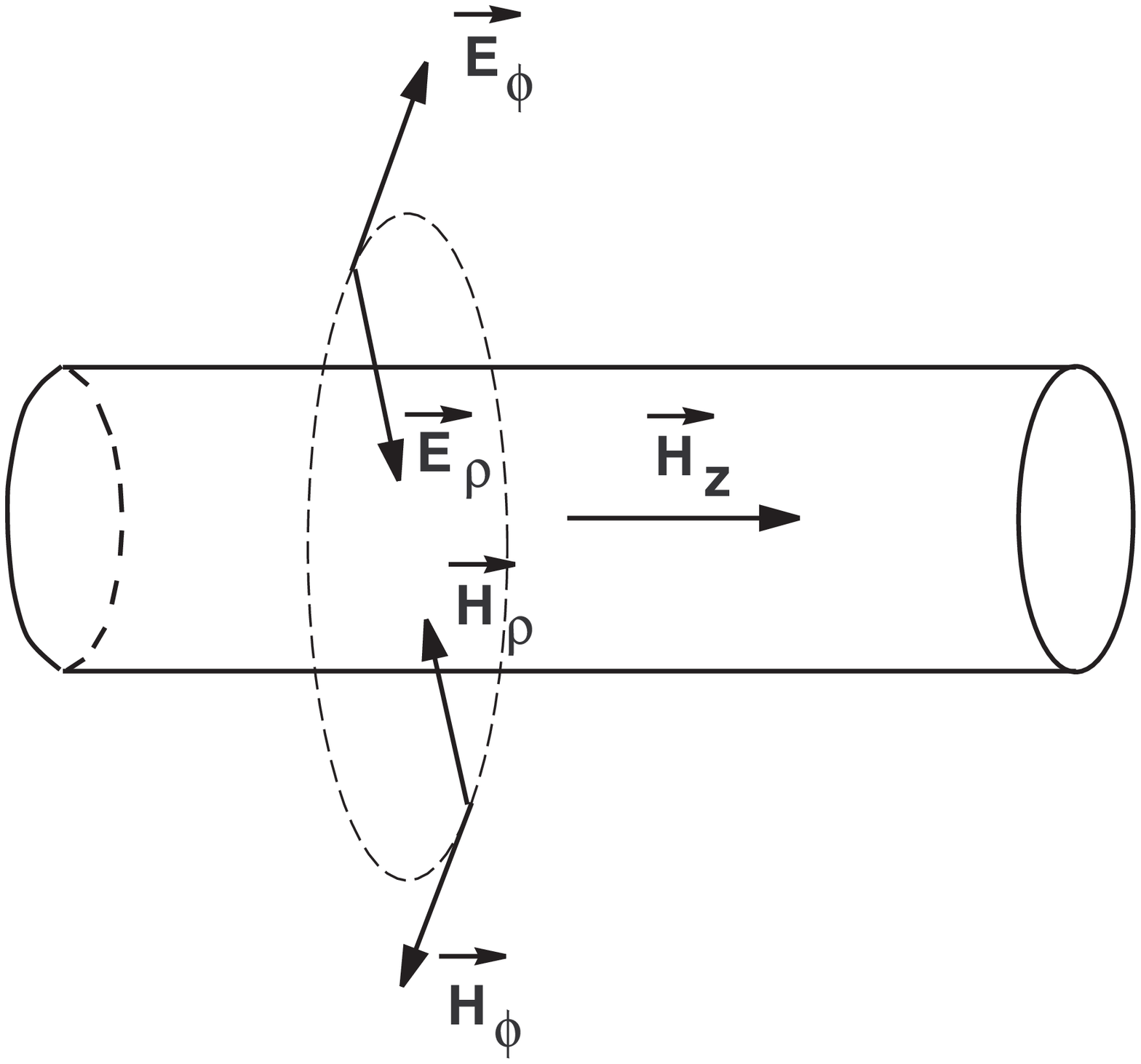}}
    \caption{The flux tube filled with the longitudinal magnetic 
    $H^2_z$ and dressed with color electric $E^a_{\rho,\varphi}$ and magnetic 
    $H^a_{\rho,\varphi}$ fields.}
    \label{fig4}
  \end{center}
  \end{minipage}
\end{figure}

The linear energy density $\varepsilon(\rho)$ for this tube is 
\begin{equation}
\begin{split}
    2\varepsilon(\rho) = {f'}^2 + {v'}^2 + \frac{{w'}^2}{x^2} + {\phi'}^2 + 
    \frac{f^2 w^2}{\rho^2} + f^2 \phi^2 + \frac{v^2 w^2}{\rho^2} + \\
    \frac{w^2 \phi^2}{\rho^2} + \frac{\lambda}{2} 
    \left(
    \phi^2 - \phi^2_\infty
    \right)^2 + 
    \left(
    f^2 + v^2
    \right) m^2
\label{sec1-210}
\end{split}
\end{equation}
and it is presented on Fig.\ref{fig5}.

\section{Quantum interpretation}

In this section we would like to present certain arguments that the derived 
above dressed flux tube can be considered as a \textit{pure quantum} object 
in the SU(3) quantum theory on the nonperturbative level in the first approximation.
\par 
One can suppose that in QCD there are two different degrees of freedom 
\cite{vdsin1}, \cite{vdsin2}: 
the first are nonperturbative and the second perturbative degrees of freedom. 
For us is interesting here only nonperturbative degrees of freedom. In Ref's 
\cite{vdsin1}, \cite{vdsin2} it was supposed that they can be splitted on 
ordered and disordered phases: 
\begin{enumerate}
\item The gauge field components $A^a_\mu \in SU(2), a=1,2,3$ 
      belonging to the small subgroup $SU(2) \subset SU(3)$ are in an ordered phase. 
      It means that 
\begin{equation}
  \left\langle A^a_\mu (x) \right\rangle  =(A^a _{\mu} (x))_{cl}.
\label{sec4-10}
\end{equation}
      The subscript means that this is the classical field. It is assumed 
      that in the first approximation these degrees of freedom are classical 
      and is described by appropriate classical equations. 
\item The gauge field components $A^m_\mu$ (m=4,5, ... , 8) and 
      $A^m_\mu \in SU(3)/SU(2)$) belonging to the coset SU(3)/SU(2) are in 
      a disordered phase (or in other words - a condensate), but have 
      non-zero energy. It means that 
\begin{equation}
  \left\langle A^m_\mu (x) \right\rangle = 0, 
  \quad \text{but} \quad 
  \left\langle A^m_\mu (x) A^n_\nu (x) \right\rangle \neq 0 .
\label{sec4-20}
\end{equation}
      These degrees of freedom are pure quantum degrees and are involved in the 
      equations for the ordered phase as an averaged field distribution of coset 
      components. 
\end{enumerate}
In Ref.\cite{vdsin2} was made the following assumptions and simplifications:
\begin{enumerate}
\item 
    The correlation between coset components $A^m_\mu (y)$ and $A^n_\nu (x)$ 
    in two points $x^\mu$ and $y^\mu$ is 
    \begin{eqnarray}
        \left\langle A^m_\mu (y) A^n_\nu (x) \right\rangle &=&        = 
        - \eta_{\mu\nu} \mathcal{G}^{mn} (y,x)
    \label{sec2-11}\\
        \mathcal{G}^{mn} (y,x) &=& - \frac{1}{3}f^{mpb} f^{npc} \phi^b (y) \phi^c (x) 
    \label{sec2-11a}    
    \end{eqnarray}
    where $f^{abc}$ is the structure constants of the SU(3) group. 
\item There is not correlation between 
      ordered (classical) and disordered (quantum) phases 
    \begin{equation}
      \left\langle f(a^a_\mu) g(A^m_\nu) \right\rangle =
      f(a^a_\mu)  \left\langle g(A^m_\mu) \right\rangle
    \label{sec2-13}
    \end{equation}
    where $f$ and $g$ are arbitrary functions. 
\item 
    The 4-point Green's function is 
    \begin{equation}
        \left\langle
        A^m_\alpha (x) A^n_\beta (y) A^p_\mu (z) A^q_\nu (u) 
        \right\rangle = 
        \left(
        E^{mnpq}_{1,abcd} \eta_{\alpha\beta} \eta_{\mu\nu} + 
        E^{mpnq}_{2,abcd} \eta_{\alpha\mu} \eta_{\beta\nu} + 
        E^{mqnp}_{3,abcd} \eta_{\alpha\nu} \eta_{\beta\mu}
        \right) 
        \phi^a (x) \phi^b(y) \phi^c (z) \phi^d(u) 
\label{sec2-15}
\end{equation}
\end{enumerate} 
here $E^{mnpq}_{1,abcd}, E^{mpnq}_{2,abcd}, E^{mqnp}_{3,abcd}$ are constants. 
The main idea proposed in Ref.\cite{vdsin1} is that the initial SU(3) Lagrangian 
\begin{equation}
    \mathcal{L}_{SU(3)} = -\frac{1}{4}F^A_{\mu\nu} F^{A\mu\nu} \; 
    A = 1,2, \cdots 8 .
\label{sec2-10}
\end{equation}
after above mentioned assumptions and simplifications can be reduced to the 
SU(2) Yang - Mills - Higgs Lagrangian 
\begin{equation}
  \mathcal{L}_{SU(2)} = - \frac{1}{4}  F^a_{\mu\nu} F^{a\mu\nu} + 
  \frac{1}{2}   \left(
  \partial_\mu \phi^a - \frac{g}{2} \epsilon^{abc} A^b_\mu \phi^c
  \right)^2 + \frac{m^2_\phi}{2} (\phi ^a \phi ^a ) 
  - \lambda \left( \phi^a \phi^a \right)^2  +
  \frac{g^2}{2} a_{\mu} ^b \phi ^b a^{c \mu} \phi ^c .
\label{sec2-20}
\end{equation}
The first term $F^a_{\mu\nu} F^{a\mu\nu}$ is the Lagrangian for the 
ordered phase $A^a_\mu$ and the next terms are the Lagrangian for the disordered 
phase (for the condensate). There is an additional gauge noninvariant term 
$\frac{g^2}{2} a_{\mu} ^b \phi ^b a^{c \mu} \phi ^c$. For the ans\"atz 
\eqref{sec1-30} the corresponding terms in field equations are zero and 
it is nonimportant for us. 
\par 
The next remark concerns to the mass $m(a)$. In during of derivation Lagrangian 
\eqref{sec2-20} it was assumed that there is a mechanism of generating 
the mass term $m^2_\phi \phi^a \phi^a$ for the disordered phase. By the same way 
one can suppose that an identical mechanism works for the ordered phase 
$A^a_\mu$
\begin{equation}\label{sec2-30}
    m^2(a) A^a_\mu A^{a \mu} = 
    m^2(1) A^a_\mu A^{a\mu} + m^2(2) A^a_\mu A^{a\mu} + 
    m^2(3) A^a_\mu A^{a\mu} .
\end{equation}
This term will appear in Lagrangian \eqref{sec2-20} and in field equations 
\eqref{sec1-10}, additionally we suppose that $m(1)=m(2)=m$ and $m(3)=0$. 
Therefore we see that above presented dressed flux tube is 
\textit{a pure quantum flux tube} in SU(3) quantum theory.

\section{Concusions}

In this paper we have found the flux tube solution in the SU(2) 
Yang - Mills - Higgs theory with broken gauge symmetry: it is the 
Nielsen - Olesen flux tube dressed with the radial and azimuthal color electric 
and magnetic fields. It was shown that this dressed flux tube can be considered 
as the tube on the nonperturbative level in quantum SU(3) gauge theory.
\par 
It is necessary to note that this tube solution exists only in the presence of 
gauge symmetry breakdown: $m(1,2) \neq 0$ and $m_\phi \neq 0$. It means that 
there is very close relation between confinement and symmetry breaking 
phenomena: they are tied in a tight knot. Currently there are different mechanism 
for generating mass term: the condensation of ghost fields \cite{dudal}, 
the Coleman - Weinberg mechanism \cite{coleman1}, the nonperturbative 
symmetry breaking mechanism \cite{dzhun}. Nevertheless it is not clear how 
can these mechanisms applicable in the situation presented here. One can say 
that this problem is not simpler confinement problem and is the task for the 
future \textit{nonperturbative} investigations. 
\par 
Another interesting feature is that the dressed tube exists only by certain 
values of mass $m$. Probably it is connected with the fact that in full SU(3) 
quantum theory the mass $m$ will be explicitly calculated that gives us the 
mass $m$. In our situation the presented here value $m^2 \approx 0.9402$ 
is an approximation to a precise value. 
\par 
We have to note that the presented here the nonperturbative technique strongly 
differs from perturbative calculations based on Feynman diagrams. Following 
to Heisenberg \cite{heis} the nonperturbative calculations in quantum field 
theory are based on an infinite equations set which connect all Green's 
functions. In fact the initial equations \eqref{sec1-10}, \eqref{sec1-20} 
are some approximations to this infinite equations set. Our approximation 
with the ordered and disordered phases allows us to reduce the infinite 
equations set to the finite field equations \eqref{sec1-10}, \eqref{sec1-20}. 
The Heisenberg quantization idea is similar to a turbulence theory where 
all quantities (velocities, pressure and so on) are correlated at different 
points. These correlators (Green's functions for quantum field theory) are 
nor zero for $t=const$ (in contrast to propagators in quantum field theory). 
By the same way in quantum field theory with strong interactions Green's 
functions will be nonzero at one moment 
$\left\langle Q \left| A^a_\mu(t,\vec{x}) 
A^b_\mu(t,\vec{y}) \right| Q \right\rangle \neq 0$ in contrast with the case 
of linear fields where the propagator is zero for one moment. It means that 
in linear quantum theories (or in theories with weak interactions) the 
interaction is carried by quanta but in quantum field theory with strong 
interaction (like to turbulence) can exist a field distributions where the 
field magnitudes at the different points and at one moment are correlated.

\section{Acknowledgments}
I am very grateful to the ISTC grant KR-677 for the financial support.

\end{document}